\documentclass[a4paper,british]{article}
\usepackage[T1]{fontenc}
\usepackage[latin1]{inputenc}
\usepackage{amssymb}
\usepackage{float}
\usepackage{graphicx}
\usepackage{setspace}

\usepackage{color}

\makeatletter

\usepackage{babel}
\makeatother
\begin{document}

\begin{center}{\Large ExHuME: A Monte Carlo Event Generator for Exclusive Diffraction.}
\end{center}{\Large \par}

\begin{singlespace}
\begin{center}\textbf{J. Monk and A. Pilkington}\\
\emph{School of Physics and Astronomy, University of Manchester}\\
\emph{Manchester M13 9PL, England}\\
\end{center}
\end{singlespace}

\begin{abstract}
We have written the Exclusive Hadronic Monte Carlo Event (ExHuME) generator.  ExHuME is based around the perturbative QCD calculation of Khoze, Martin and Ryskin of the process $pp\rightarrow p+X+p$, where $X$ is a centrally produced colour singlet system.  
\end{abstract}

\section*{Introduction}

\begin{figure}[H]
\begin{center}
\includegraphics[%
  bb=20bp 465bp 573bp 760bp,
  clip,
  width=0.50\paperwidth,
  keepaspectratio]{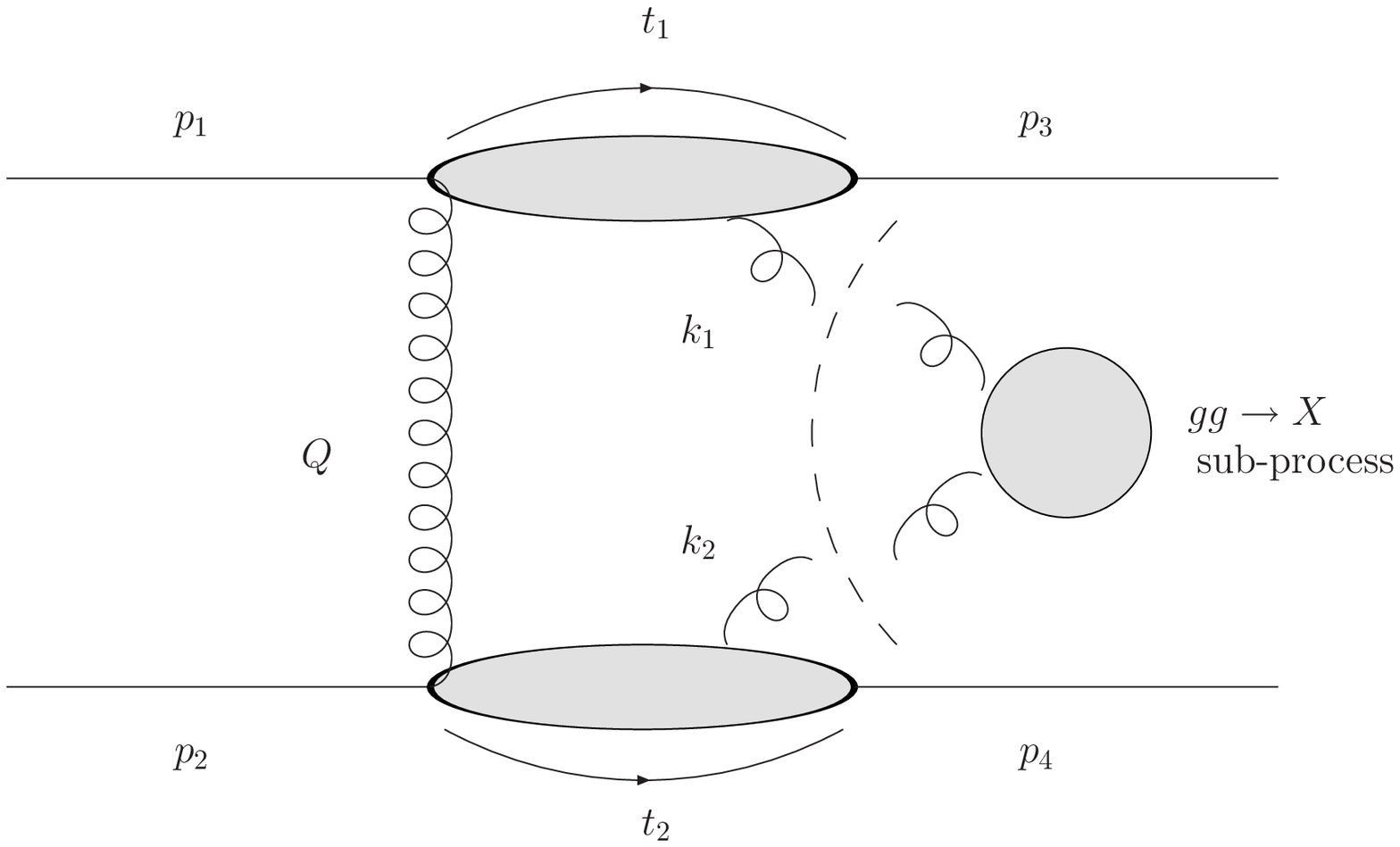}
  \end{center}
\caption{\label{cap:excldiff}The Exclusive Diffractive Process}
\end{figure}

Central exclusive production events are those in which the colliding
protons remain intact after an interaction and all of the energy transferred
from the protons goes into a central colour singlet system. Unlike inclusive double pomeron exchange, in which pomeron remnants are present, there is no radiation other than that from the central system. Furthermore, in the limit that the transverse momentum of each of the outgoing protons vanishes, no angular momentum is exchanged between the proton lines and the quantum numbers of the central system are constrained to be $J_{Z}=0$ with even parity.  The leading order diagram for the central exclusive production of  system $X$ is shown in figure \ref{cap:excldiff}.

The invariant mass of the central system can be determined if proton detectors mounted very forward of the central detector are used to measure the longitudinal momentum losses, $x_{1,2}$, and momentum transfers squared, $t_{1,2}$. 

ExHuME factorises central exclusive production as indicated by the dashed line in figure \ref{cap:excldiff}.  We use gluon fusion cross sections - taking the colour singlet and spin zero projections
of the amplitude \cite{prospects} -  but do not fix the collision energy. The luminosity, $\mathcal{L}$, for the fusing gluons is dependent on both the invariant mass ($\sqrt{\hat{s}}$) and rapidity, ($y=\frac{1}{2}\log(x_{1}/x_{2})$) of the central system \cite{prospects, higgsrapgap, higgsseen?}:

\begin{equation}
\hat{s}\frac{\partial\mathcal{L}}{\partial\hat{s}\partial y\partial t_{1}\partial t_{2}}=
e^{b(t_{1}+t_{2})}\left(\frac{\pi}{8}\int \frac{dQ_{\perp}^{2}}{Q_{\perp}^{4}}f_{g}(Q_{\perp}, \mu, x_{1}, x_{1}^{\prime})f_{g}(Q_{\perp}, \mu, x_{2}, x_{2}^{\prime})\right)^{2}.\label{eq:lumi}
\end{equation}
The forward limit implies that $k_{1\perp}\simeq -k_{2\perp}\simeq-Q_{\perp}$,
which simplifies the integral to the form in equation (\ref{eq:lumi}). The $f_{g}$ are the  unintegrated skewed
gluon distributions (which can be calculated from the integrated gluons \cite{PhysicsTaggedForward}).  We choose $\mu=0.618\sqrt{\hat{s}}$ \cite{mu} for the hard scale and $b=4$ GeV$^{-2}$ \cite{prospects}.  The $x^{\prime}$ are the longitudinal momentum fractions of the gluon that is not connected to the central system, $x_{1}^{\prime}=x_{2}^{\prime}= \mathcal{O}(Q_{\perp}^{2}/s)$ for symmetric proton beams of energy $\sqrt{s}/2$.
The integral in  equation (\ref{eq:lumi}) converges in the infra-red due to the presence
of a Sudakov factor that imposes the requirement that there be no
radiation emitted between the scales of $Q_{\perp}$ and $\mu$. The leading term
in $k_{1\perp}\cdot k_{2\perp}$ in the integrand of equation (\ref{eq:lumi}) is $Q_{\perp}^{2}$ and has no dependence on the azimuthal angle $\phi$ between
the outgoing proton directions. For complete details of the luminosity
calculation we refer to the calculation of V. A. Khoze et al \cite{prospects, higgsrapgap, higgsseen?, PhysicsTaggedForward, mu}, upon which ExHuME is based.

\begin{figure}[H]
\begin{center}

\includegraphics[%
bb=85bp 10bp 565bp 720bp,
  clip,
width=0.5\textwidth,
keepaspectratio,
angle=270]{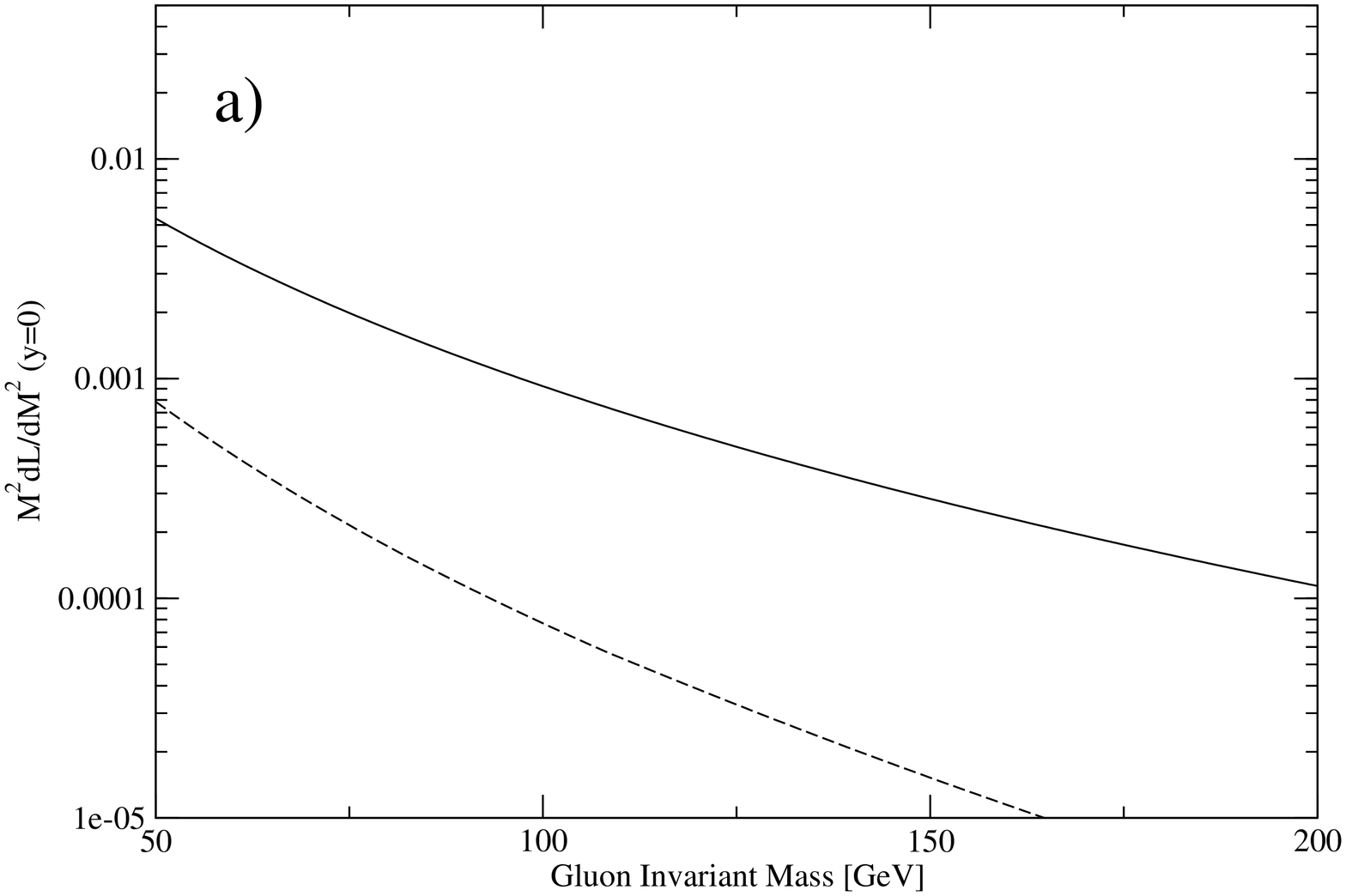}
\\*
\includegraphics[%
bb=85bp 10bp 565bp 720bp,
  clip,
width=0.5\textwidth,
keepaspectratio,
angle=270]{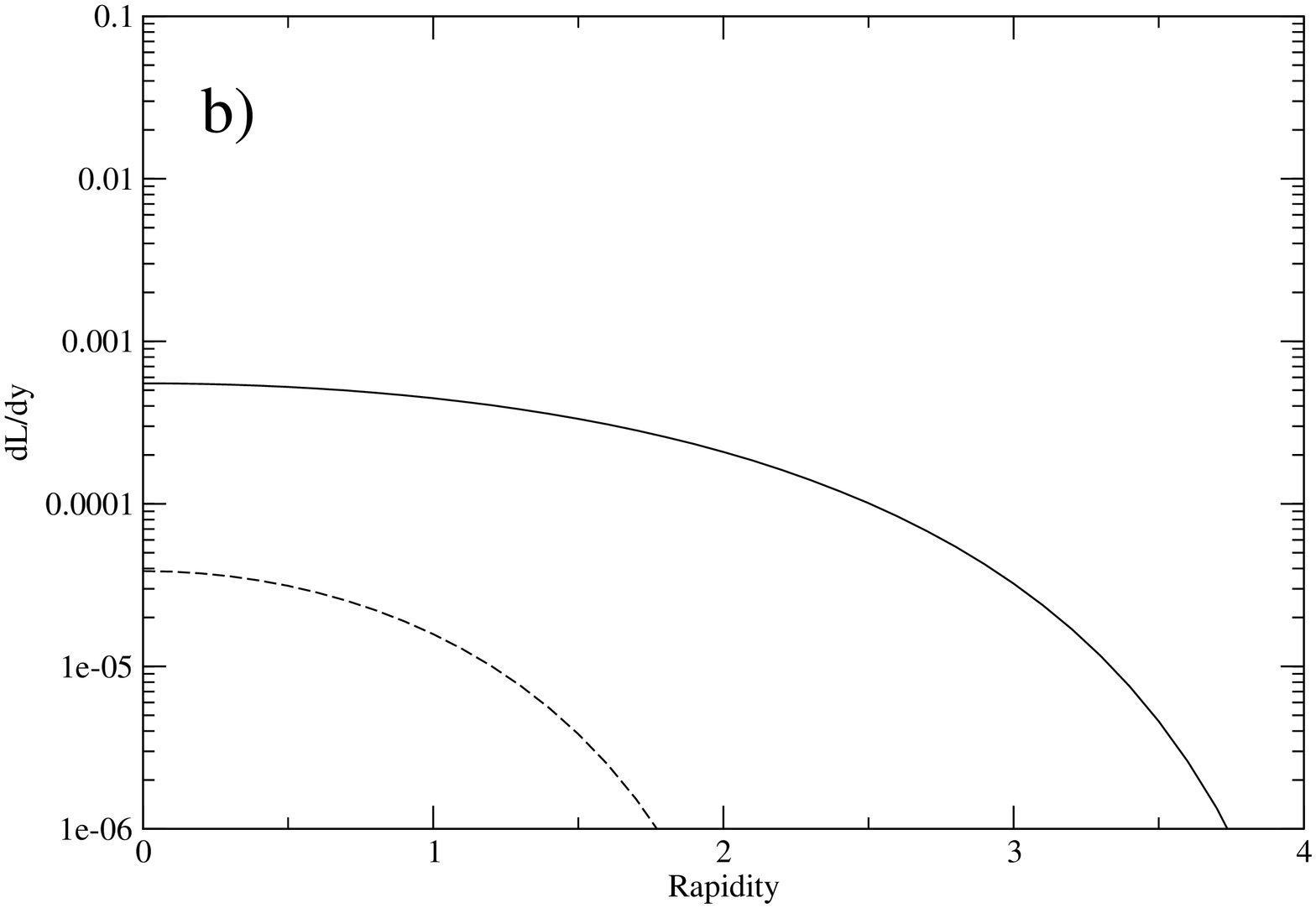}

\end{center}

\caption{\label{cap:lumi}The default luminosity used in ExHuME at collider energies of 1.96 TeV (dashed line) and 14 TeV (solid line).  a)  shows the luminosity vs. the invariant mass of the fusing gluons at zero rapidity.  b) shows the rapidity dependence of the luminosity when the invariant mass of the fusing gluons is fixed at 120 GeV.  A fixed survival factor of $0.03$ ($14$ TeV) or $0.045$ ($1.96$ TeV) has been included and the luminosity has been integrated over the $p_{\perp}$ of the outgoing protons.}
\end{figure}

A survival factor, $\mathcal{S}^{2}$, is also required to
ensure that there are no additional interactions between the proton
lines. We use a constant value for $\mathcal{S}^{2}$ that is fixed for a given collision energy, although we anticipate that further work will lead to a parameterisation of the survival factor in terms of $t_{1}$, $t_{2}$ and $\phi$ \cite{PhysicsTaggedForward}.  We note that the $\phi$ dependence in that model is approximately flat for values of $p_{3\perp}, p_{4\perp}\lesssim200$MeV.  The default differential luminosity curves used by ExHuME are shown in figure \ref{cap:lumi}.

In this initial release of ExHuME we provide the following sub-processes:  $gg\rightarrow H$, $gg\rightarrow Q\bar{Q}$ and $gg\rightarrow gg$, where $H$ is a Standard Model Higgs and $Q$ is a massive quark.  We require a width for resonance production and use the Higgs line shape described in \cite{HiggsLineShape} with the Higgs width and branching ratios calculated using the program Hdecay \cite{hdecay}

\section*{Design}

ExHuME is written in a modular way using C++.  There are two main classes that are needed to generate central exclusive events.  The first is a  \texttt{CrossSection} class that calculates the differential luminosity, the gluon fusion sub-process and the kinematics of any outgoing particles.  The second class needed is an \texttt{Event} class that  generates the events.  In addition there is a \texttt{Weight} class that provides random numbers distributed according  to any given function and from which the \texttt{Event} class inherits.

\subsection*{The CrossSection Class}

The \texttt{CrossSection} class exploits the factorisation of figure \ref{cap:excldiff}.  \texttt{CrossSection} is an abstract base class containing the calculation of the effective luminosity of the gluon-gluon collision with a virtual method for the gluon fusion sub-process. Complete processes are created by inheriting from the \texttt{CrossSection} and explicitly defining a sub-process. This makes it relatively simple to implement new sub-processes in addition to the currently implemented Standard Model Higgs, $gg$ and $Q\bar{Q}$ sub-processes.  Instructions for creating a new sub-process are posted at \cite{exhume-me} and users are invited to submit any new process to the authors for inclusion in future updates to ExHuME.

\subsubsection*{CrossSection Methods}

The constructor for the \texttt{CrossSection} class is
\\* \\* \texttt{CrossSection(int, char**)},
\\ \\ which allows the user to pass a card file from the command line that changes the default values used in the luminosity calculation (see Appendix \ref{cardfiles}).  The constructors for the derived classes are \texttt{Higgs(int, char**)}, \texttt{QQ(int, char**)}, \texttt{GG(int,char**)} or \texttt{Dummy(int, char**)}. 

The invariant mass and  rapidity of the central system and the momentum transfers and azimuthal angles, $\phi_{1,2}$, of the outgoing protons can be set by the \texttt{CrossSection} method
\\ \\ \texttt{void SetKinematics(\\*
\indent const double \&mass, const double \&rapidity, \\*
\indent const double \&t1, const double \&t2,\\*
\indent const double \&phi1, const double \&phi2)};.\*
\\*
\\The method\*
\\ \\ \texttt{double Differential()}
\\ \\ then returns the differential cross section.  The 4-vectors of the outgoing protons can be accessed by the methods
\\ \\ \texttt{HepLorentzVector GetProton1()} and
\\* \texttt{HepLorentzVector GetProton2()}.
\\ \\ The sub-process level information can be extracted by the method
\\ \\ \texttt{std::vector$<$Particle$>$ GetPartons()} 
\\ \\ which returns the outgoing particles from the gluon fusion matrix element prior to decay, parton showering and hadronisation. The \texttt{Particle} class that \texttt{GetPartons()} returns contains the particle momentum, \texttt{p}, the PDG ID code, \texttt{id} and vertex, \texttt{vtx}.  It is also possible to access the kinematics of the cross section by the following methods:
\\ \\ \texttt{double GetRoot\_s()};\\* Returns the invariant mass of the colliding beams.
\\ \\ \texttt{double GetsHat()};\\* Returns the invariant mass squared of the central system.
\\ \\ \texttt{double GetSqrtsHat()};\\* Returns the invariant mass of the central system.
\\ \\ \texttt{double Getx1()}; \\* Gets the value of $x_{1}$.
\\ \\ \texttt{double Getx2()}; \\* Gets the value of $x_{2}$. 
\\ \\ \texttt{double Gett1()}; \\* Gets the value of $t_{1}$. 
\\ \\ \texttt{double Gett2()}; \\* Gets the value of $t_{2}$. 
\\ \\ \texttt{double GetPhi1()}; \\* Gets the value of $\phi_{1}$, the azimuthal angle between the outgoing proton 1 and the $x$ direction.
\\ \\ \texttt{double GetPhi2()}; \\* Gets the value of $\phi_{2}$, the azimuthal angle between the outgoing proton 2 and the $x$ direction.
\\ \\ \texttt{double GetEta()}; \\* Gets the rapidity, $y$,  of the central system.
\\ \\ Hadronisation is performed via the method
\\ \\ \texttt{void Hadronise()}
\\ \\ which places the event record into an external \texttt{hepevt} common block that can be written to file.

There are a number of sub-process specific methods. The Higgs decay type can be set by the method
\\* \\* \texttt{SetHiggsDecay(const int\&)}
\\ \\ with the PDG code of the decay products as the argument. Similarly the quark type in $q\bar{q}$ can be set by the method
\\* \\* \texttt{SetQuarkType(const int\&)}
\\ \\ which defaults to $b\bar{b}$ production. In addition, the \texttt{GG} and \texttt{QQ} classes contain the method
\\* \\* \texttt{SetThetaMin(const double\&)}
\\ \\ which sets the minimum (and maximum) polar angle of an outgoing parton relative to the beamline in the rest frame of the central system. The default value is $\cos\theta = 0.95$. 

\subsection*{The Event Class}

The \texttt{Event} class generates events, calculates the total cross section and reports the efficiency with which events were generated once event generation has finished.  \texttt{Event} contains a pointer to a \texttt{CrossSection} and the user specifies which sub-process is to be generated in the constructor for the \texttt{Event}. The \texttt{Event} class also inherits from the \texttt{Weight} class. The numerical weighting algorithm is initialised to return the mass distribution of the differential cross section at a rapidity zero and is effective even for a narrow resonance such as the Higgs.  The variables $t_{1}$ and $t_{2}$ are distributed according to $e^{b(t_{1} + t_{2})}$ whilst $\phi_{1}$, $\phi_{2}$ and $y$ are uniformly distributed.  

\subsubsection*{Event Methods}

The event is defined by calling the constructor \\ \\ \texttt{Event(CrossSection\& P, const unsigned int R)}, \\* \\* where P is a \texttt{CrossSection} with the sub-process defined and R is a random number seed. There are a number of methods that can be used to set the kinematic ranges of the parameters used to define the event: 
\\ \\ \texttt{void Setx1Max(const double\&)}; \\*  Sets the upper limit of $x_{1}$.
\\ \\ \texttt{void Setx2Max(const double\&)}; \\* Sets the upper limit of $x_{2}$.
\\ \\ \texttt{void Sett1Max(const double\&)}; \\*  Sets the upper limit of $t_{1}$ ($t\le0$).
\\ \\ \texttt{void Sett1Min(const double\&)}; \\* Sets the lower limit of $t_{1}$.
\\ \\ \texttt{void Sett2Max(const double\&)}; \\*  Sets the upper limit of $t_{2}$.
\\ \\ \texttt{void Sett2Min(const double\&)}; \\* Sets the lower limit of $t_{2}$.
\\ \\ \texttt{void SetMassRange(const double \& minimum, const double \& maximum)}; \\*  Sets the lower and upper limits of the mass range respectively. 
\\ \\ The last method that is used before event generation is \\ \\ \texttt{void SetParameterSpace()}, \\* \\*  which must be called in order to initialise the event generation. Individual events are generated by the method
\\ \\ \texttt{void Generate()}.
\\ \\ Finally, the total cross section from the events generated can be calculated by the method
\\ \\ \texttt{double CrossSectionCalculation()} 
\\ \\and the efficiency via
\\ \\ \texttt{double GetEfficiency()}.

\section*{Using ExHuME}

\subsection*{Installing ExHuME}

The ExHuME source code is available from \cite{exhume-me} or on request from the authors.  In its standard form ExHuME must be linked at compilation to Pythia \cite{pythia}, CLHEP \cite{clhep} and either LHAPDF \cite{lhapdf} or the CERN PDFLIB \cite{pdflib}.  It would also be possible to modify ExHuME to use Herwig \cite{herwig} instead of Pythia for the hadronisation or to use a stand alone PDF instead of either LHAPDF or PDFLIB.  By default ExHuME sets the location of the directory containing the grid or parameter files for LHAPDF to be wherever the program is executed from.  A symbolic link should be created to wherever the grid and parameter files actually reside.  For further information please see the respective documentation for each of these programs .

\subsection*{Example Main Program}

In this section we demonstrate a simple main program that generates 5000 $H \rightarrow WW^{*}$ events for a Higgs with the default mass of 120 GeV. We also show how to extract simple information from the hepevt record.
\\
\\ \texttt{\indent \#include <iostream>}
\\ *
\\The following headers are for ExHuME:
\\
\\* \texttt{\indent \#include "Event.h"}
\\* \texttt{\indent \#include "Higgs.h"}
\\*
\\* \texttt{int main(int argc, char** argv)\{}
\\*
\\ Declare a new Higgs \texttt{CrossSection}:
\\
\\* \texttt{\indent Exhume::Higgs higgs(argc,argv);}
\\
\\ and set the Higgs to decay only to W bosons:
\\
\\* \texttt{\indent higgs.SetHiggsDecay(24); // 24 is the PDG code for W}
\\
\\ Declare an event with the Higgs as the cross section and a random number seed of 1111.
\\
\\* \texttt{\indent Exhume::Event HiggsEvent(higgs,1111);}
\\
\\* The allowed range of gluon fusion invariant masses must be set.  As long as the range is much bigger than the width of the resonance  the results will not be sensitive to the range chosen.  This is, of course, not the case whenever the central system does not have a narrow width, for example in $gg$ and $Q\bar{Q}$ production.  The efficiency of event generation will drop if a large mass range is chosen for a narrow resonance such as the Higgs, so for this example where the width is 0.0036 GeV we set the mass range to be between 115 GeV and 125 GeV.
\\
\\* \texttt{\indent HiggsEvent.SetMassRange(115,125);   }
\\
\\* This must be called before event generation can begin:
\\
\\* \texttt{\indent HiggsEvent.SetParameterSpace(); }
\\
\\* \texttt{\indent double x1; \\* \indent int Nobj; \\* \indent std::vector$<$Exhume::Particle$>$ HiggsInfo;}
\\
\\* \texttt{\indent for ( int i $=$ 0 ; i $<$ 5000 ; i$++$ )\{ } 
\\
\\* The next line generates a single event:
\\
\\* \texttt{\indent \indent HiggsEvent.Generate();}
\\
\\* Access the longitudinal momentum loss of proton 1 for this event:
\\
\\* \texttt{\indent \indent x1 $=$ process.Getx1();}
\\
\\* and the information about the Higgs:
\\
\\* \texttt{\indent \indent HiggsInfo $=$ process.GetPartons();} 
\\
\\* Get the number of particles in the hepevt common block:
\\
\\* \texttt{\indent \indent Nobj $=$ hepevt\_.nhep;}  
\\* \texttt{\indent \} }
\\* \texttt{\indent std::cout$<<$"   Cross section =  "\\* \indent $<<$HiggsEvent.CrossSectionCalculation()$<<$std::endl;}
\\* \texttt{\indent std::cout$<<$"   Efficiency of event generation = "\\* \indent $<<$HiggsEvent.GetEfficiency()$<<$std::endl;}
\\* \texttt{\indent return(0);}
\\* \texttt{\}}
\\
\\ The program allows (but does not demand) a card file to be given on the command line that overrides the default parameters. Such a card file could look like
\\
\\* \texttt{HiggsMass \indent 140}
\\* \texttt{TopMass \indent 180}
\\
\\ which would be appropriate for investigating the effects of varying the higgs and top masses. 
\\
\section*{Acknowledgements}

We would like to thank Brian Cox, Jeff Forshaw, Valery Khoze and Misha Ryskin for many useful and interesting discussions and suggestions.  We also thank the U.K. Particle Physics and Astronomy Research Council for funding this work.

\appendix

\section{The ExHuME Control Parameters}\label{cardfiles}

ExHuME can be controlled by passing a card file from the command line that can contain the control parameters given in table \ref{cards}. The collider can be set to the LHC ($1$), the Tevatron ($0$) or neither ($-1$). Choosing a collider sets the proton collision energy, $\sqrt{s}$, the survival factor $\mathcal{S}^{2}$ and $R_{g}$, which accounts for the skewed effect in the un-integrated gluons. Choosing neither means that the user must set these parameters. The PDF values are the PDF set numbers accepted by the LHAPDF library.  \texttt{Freeze} is a scale below which $\alpha_{s}$ is frozen and $Q_{\perp min}$ is the lower bound on the integral in equation (\ref{eq:lumi}) - the integral is formally convergent, but the necessary freezing of $\alpha_{s}$ requires a lower bound to be introduced.

\begin{table}[htdp]
\begin{center}
\begin{tabular}{| l | l | l | l |}%these are l's in the middle
\hline
\textbf{Parameter}&\textbf{Name}&\textbf{Type}&\textbf{Default}\\
\hline
$\alpha$&\texttt{AlphaEW}&\texttt{double}& 0.0072974\\
$M_{W}$&\texttt{WMass}&\texttt{double}& 80.33\\
$M_{Z}$&\texttt{ZMass}&\texttt{double}& 91.127\\
$M_{H}$&\texttt{HiggsMass}&\texttt{double}&120.0 \\
$M_{t}$&\texttt{TopMass}&\texttt{double}&175.0 \\
$M_{b}$&\texttt{BottomMass}&\texttt{double}& 4.6\\
$M_{c}$&\texttt{CharmMass}&\texttt{double}& 1.42\\
$M_{s}$&\texttt{StrangeMass}&\texttt{double}& 0.19\\
$M_{\tau}$&\texttt{TauMass}&\texttt{double}& 1.77\\
$M_{\mu}$&\texttt{MuonMass}&\texttt{double}& 0.1057\\
$v$&\texttt{HiggsVev}&\texttt{double}& 246.0\\
$Q_{\perp min}^{2}$&\texttt{MinQt2}&\texttt{double}& 0.64\\
$\Lambda_{QCD}$ ($MeV$)&\texttt{LambdaQCD}&\texttt{double}& 80 ($MeV$)\\
&\texttt{Freeze}&\texttt{double}& $Q_{\perp min}$\\
$b$&\texttt{B}&\texttt{double}& 4.0 \\
collider&\texttt{FNAL\_or\_LHC}&\texttt{int}& 1\\
$s$&\texttt{s}&\texttt{double}& $1.96\times 10^{8}$\\
$s^{\frac{1}{2}}$&\texttt{root\_s}&\texttt{double}& 14000.0\\
$R_{g}$&\texttt{Rg}&\texttt{double}& 1.2\\
$\mathcal{S}^{2}$&\texttt{Survive}&\texttt{double}& 0.03\\
$PDF$&\texttt{PDF}&\texttt{int}& 20250\\
\hline

\end{tabular}
\end{center}
\caption{ExHuME control parameters}
\label{cards}
\end{table}

\end{document}